%% file: main.tex
\documentclass[sigconf]{acmart}

\usepackage{booktabs} 
\usepackage{subfigure}
\usepackage{balance}

\begin{document}

\copyrightyear{2017}
\acmYear{2017}
\setcopyright{acmcopyright}
\acmConference{DH '17}{July 02-05, 2017}{London, United Kingdom}\acmPrice{15.00}\acmDOI{http://dx.doi.org/10.1145/3079452.3079464}
\acmISBN{978-1-4503-5249-9/17/07}

\title{Enhancement of Epidemiological Models for\\Dengue Fever Based on Twitter Data}
 
\author{Julio Albinati, Wagner Meira Jr., Gisele L. Pappa}
\orcid{1234-5678-9012}
\affiliation{%
  \institution{Department of Computer Science\\Universidade Federal de Minas Gerais}
  \city{Belo Horizonte} 
  \state{Minas Gerais}
  \country{Brazil}
}
\email{jalbinati,meira,glpappa@dcc.ufmg.br}

\author{Mauro Teixeira, Cecilia Marques-Toledo}
\affiliation{%
  \institution{Departament of Biochemistry and Immunology\\Universidade Federal de Minas Gerais}
  \city{Belo Horizonte} 
  \state{Minas Gerais}
  \country{Brazil}
}
\email{mmtex@icb.ufmg.br, cecilia.a.marques@gmail.com}

\renewcommand{\shortauthors}{Albinati et al.}

\begin{abstract}
Epidemiological early warning systems for dengue fever rely on up-to-date epidemiological data to forecast future incidence. However, epidemiological data typically requires time to be available, due to the application of time-consuming laboratorial tests. This implies that epidemiological models need to issue predictions with larger antecedence, making their task even more difficult. On the other hand, online platforms, such as Twitter or Google, allow us to obtain samples of users' interaction in near real-time and can be used as sensors to monitor current incidence. In this work, we propose a framework to exploit online data sources to mitigate the lack of up-to-date epidemiological data by obtaining estimates of current incidence, which are then explored by traditional epidemiological models. We show that the proposed framework obtains more accurate predictions than alternative approaches, with statistically better results for delays greater or equal to 4 weeks.
\end{abstract}

%
%

\ccsdesc[500]{Applied computing~Health informatics}
\ccsdesc[300]{Information systems~Expert systems}
\ccsdesc[300]{Information systems~Social networks}
\ccsdesc[300]{Computing methodologies~Gaussian processes}


\keywords{dengue fever, Twitter, Gaussian processes}

\maketitle

\section{Introduction} \label{sec:introduction}

\input{introduction}

\section{Related Work} \label{sec:related}

\input{related}

\section{Data Collection}

\input{data}

\section{Incorporating online data into epidemiological models}

\input{methodology}

\section{Experimental Analysis} \label{sec:experiments}

\input{experiments}

\section{Conclusions and Future Work}

\input{conclusions}

\subsection*{Competing Interests}
The authors have declared that no competing interests exist.

\subsection*{Acknowledgements} 
We would like also to thank the authors from \citet{souza2015latent} for sharing their dataset of geolocalized dengue-related tweets, enabling this study.

This work was partially supported by CNPq, FAPEMIG, CAPES, InWeb, MASWeb, BIGSEA, INCT-Cyber and INCT-Dengue.

\balance

\bibliographystyle{ACM-Reference-Format}
\bibliography{paper}

\end{document}

%% file: introduction.tex
Dengue fever is a mosquito-borne viral disease whose pathogen is mainly transmitted by females mosquitoes of the species \emph{Aedes aegypti}, the same mosquito that carries the viruses of zika, yellow fever and chikungunya. As a severe flu-like illness, it may cause high fever, strong headache, pain behind the eyes, muscles and joints, nausea, vomiting, swollen glands and rash, which can last for one week. There is no specific treatment for dengue fever nor widespread available vaccines and, therefore, control of the disease is mainly performed by suppressing the vector population, as well as identifying outbreaks as quickly as possible \citep{who2016factsheet}. Although case fatality rate can be as low as 1\% with proper treatment, the disease comes with great economical and social burden~\citep{samir2013global}.

According to \citet{samir2013global}, dengue fever is ubiquitous throughout the tropics. The study estimates about 390 million dengue infections worldwide every year, with 96 million of these cases being symptomatic. The Americas contribute with 14\% of the symptomatic infections, of which over one fourth occurs in Brazil. This scenario motivates the development of appropriate \emph{early warning systems} (EWSs) to quickly identify new dengue fever outbreaks in Brazilian cities.  An EWS is a tool capable of quickly identifying/forecasting risks, and plays a major role on disaster risk reduction by preventing loss of life and mitigating economical and/or material impact \citep{wiltshire2006developing}. In the context of epidemiology, an epidemiological EWS (EEWS) can be a fundamental step on implementing effective interventions to control infectious diseases, reducing mortality and morbidity in human populations, since it allows authorities to plan ahead and act appropriately.

The main requirement of an EEWS is the availability of up-to-date epidemiological data, which will be used in pattern extraction for forecasting of future incidence of the disease under study. However, epidemiological data cannot be assumed to be provided in real-time. It is common for this kind of data to require time-consuming tests for confirmation of suspected cases. Besides that, the propagation of information from local health care agents to national health authorities may also require extra time. Therefore, it is more reasonable to assume that, at a given moment in time $t$, the most recent epidemiological data available is associated with time $t - \gamma$, where $\gamma$ is a positive number. This delay in the report of epidemiological data can be highly detrimental to EEWS, specially for infectious diseases, where current incidence is a valuable information for predicting near future incidence \citep{albinati2016accurate,dom2013generating,eastin2014intra}.

On the other hand, nowadays a large number of people use text-based online social networks, such as Twitter or Facebook, to discuss various subjects, including personal matters. Similarly, they use search engines and Wikipedia, looking for information. These platforms are capable of providing samples of users' interactions in real-time, leading to the consolidation of useful data sources that may be exploited in an easy and systematic manner. These online data sources have attracted the attention of many researchers who have gone on to use the data to predict real-life events \citep{broniatowski2013national,gomide2011dengue,generous2014global,sakaki2010eathquake,shaman2013real}, effectively using people's interaction with these platforms as sensors.

In this work, we combine real-time online data sources with epidemiological data to deal with this time lag required for epidemiological dengue data to be ready for use. We propose a framework that exploits data coming from Twitter to estimate delayed epidemiological dengue data, which is then used together with actual epidemiological data in an EEWS.  In our experimental analysis, we compared the proposed framework with alternative approaches and verified that the proposed framework leads to more accurate predictions and that it can be safely applied even on cities for which Twitter data is not abundant, as it automatically identifies scenarios where online data can be usefully employed to improve accuracy .

%% file: related.tex
The real-time nature of online data makes it attractive for surveillance tasks, particularly in epidemiological contexts, where many strategies that explore this kind of data to estimate incidence of various diseases around the world have been developed. For instance, in \cite{althouse2011prediction}, the authors used the number of searches on Google containing a specific set of keywords to estimate dengue fever incidence in Singapore and Bangkok, Thailand. Similarly, the authors of \cite{shaman2013real} developed a methodology that estimates influenza incidence in American cities based on data provided by Google Flu Trends, which provides estimates of influenza incidence based on search activity on Google.
  
Twitter is also a source of online data frequently exploited for epidemiological surveillance systems. The authors of \cite{broniatowski2013national} developed a classifier capable of accurately identifying tweets indicating influenza infection and showed that the volume of these tweets over time presented similar, highly correlated behavior when compared to epidemiological data. A similar idea was investigated in \cite{gomide2011dengue}, which developed a model for classifying dengue-related tweets into five classes: personal experience (indicating infection), opinion, jokes, information and governmental campaigns. The volume of dengue-related tweets, especially those containing personal experience, was shown to be highly correlated to epidemiological data from Brazil. This relationship was further studied in \cite{souza2015latent,souza2014evolutionary}, culminating in the development of models for estimating dengue fever incidence in Brazilian cities. 

These works, however, were developed for surveillance, and are not appropriate for EEWS. Although they are capable of estimating incidence of diseases in real-time (nowcast), they do not forecast \emph{future} incidence. The reason behind it is that online data is typically used as input features to (generalized) linear models. Therefore, in order to estimate future incidence, they would require \emph{future} online data. In other words, the fact that online data is obtained in real-time is not sufficient. 

In order to predict future outcomes, alternative ways of using online data in EEWS are required. In this direction, a possible approach was evaluated by \citet{generous2014global}, where access logs of selected Wikipedia articles were used to estimate a linear model for 7 diseases in 9 countries. In order to be able to estimate future incidence, they evaluated the usage of Wikipedia data in a lagged fashion. That is, epidemiological data for to time index $t$ would be associated with Wikipedia data for time index $t - \alpha$, where $\alpha$ is a positive integer. By doing so, the model would be capable of forecasting incidence with up to $\alpha$ time indices of antecedence. In the former, they evaluate values for $\alpha$ up to 4 weeks. In this article, we propose a distinct approach for designing truly predictive models. We obtain estimates of up-to-date epidemiological data, which can then be used within EEWS for forecasting future incidence.

%% file: data.tex
\paragraph{Epidemiological data}

We obtained the number of weekly confirmed dengue cases for 5566 Brazilian municipalities from January 2011 to October 2016, summing up to 286 weeks, provided by the Brazilian Ministry of Health. In order to account for different population sizes, we calculated the dengue incidence rate (DIR) as follows:
\begin{equation}
  DIR_{s,t} = cases_{s,t} * \frac{100000}{pop_s} \label{eq:dir}
\end{equation}
where $DIR_{s,t}$ is the DIR at city $s$ during week $t$, $cases_{s,t}$ is the number of confirmed cases at city $s$ during week $t$ and $pop_s$ is the population size at city $s$.

The Brazilian Ministry of Health has defined a system of three incidence levels for dengue fever: high, medium and low. High incidence level at a given area occurs when there is more than 300 dengue cases per 100 thousand inhabitants during one month. Medium incidence, on the other hand, means more than 100 dengue cases and less than 300 dengue cases per 100 thousand inhabitants in a month. Finally, low incidence occurs when there is less than 100 cases per 100 thousand inhabitants at the same month. In this work, however, we deal with weekly dengue data. For that reason, we redefined these incidence levels by dividing the number of required cases by 4. Therefore, a high incidence week at a given area implies in a DIR of at least 75, while medium incidence requires DIR between 25 and 75 and low incidence requires DIR smaller than 25.

In order to reduce the effort of evaluating the proposed framework, we consider only cities with more than 100 thousand inhabitants and that reached at least medium incidence during one week in the period under study, resulting in 213 cities. Although small when compared to the initial set of 5566 cities, these cities account for more than 65\% of the number of confirmed cases in Brazil. Therefore, cities where an accurate and reliable EEWS is fundamental for controlling dengue epidemics are included in the set of cities under study.

\paragraph{Twitter data}

Twitter messages (tweets) were collected by the authors of \citep{gomide2011dengue,souza2014evolutionary,souza2015latent} through the Twitter API, obtaining geolocalized tweets from January 2011 to October 2016 with at least one of the following keywords: \emph{dengue}, \emph{aedes} and \emph{aegypti}.

We aggregated tweets by city and by week in order to obtain time series containing the number of dengue-related tweets posted during each week and at each city under study.

%% file: methodology.tex
\begin{figure*}[!t]
  \centering
  \subfigure{\includegraphics[width=0.8\linewidth]{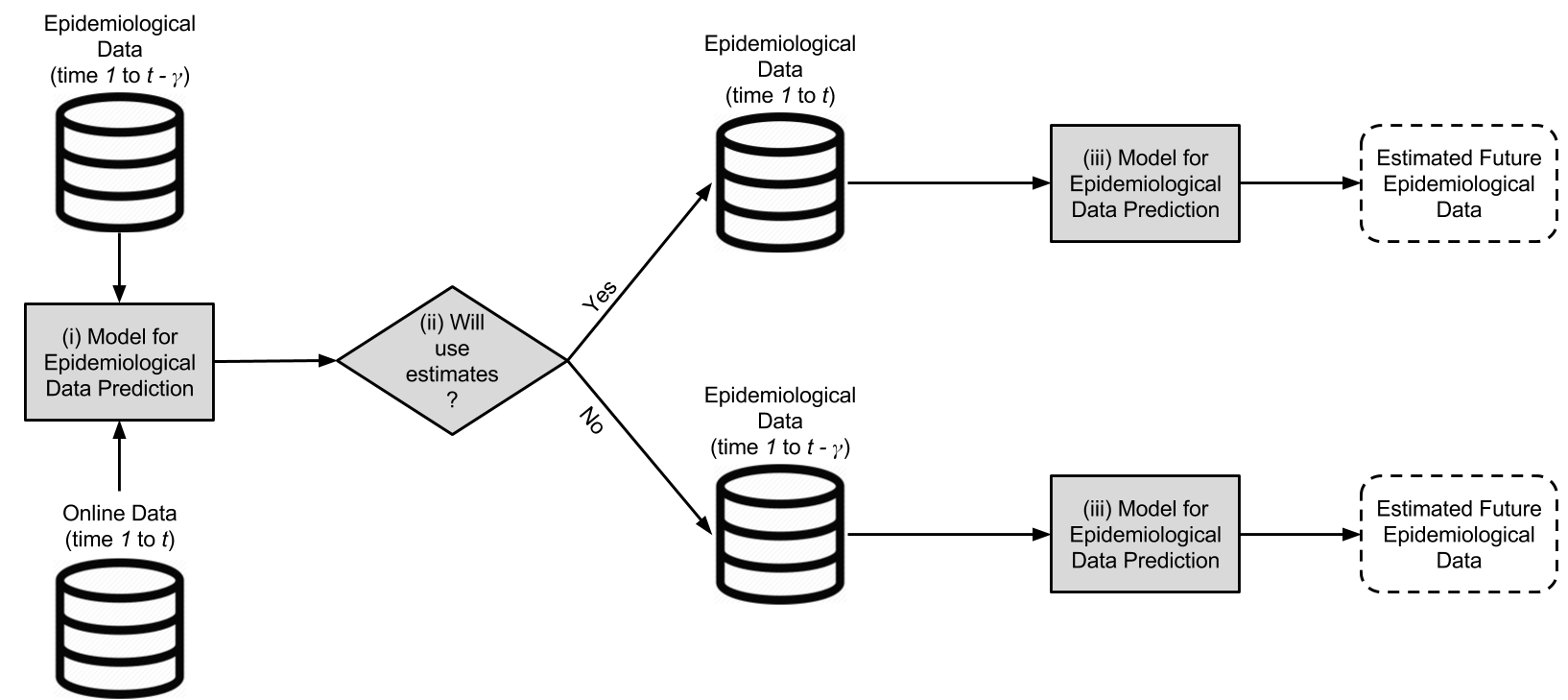}}
  \caption{The proposed framework for incorporating Twitter data into an EEWS model.}
  \label{fig:framework}
\end{figure*}

As previously indicated, in most circumstances it is not reasonable to assume that epidemiological data is provided in real time. For this reason, a prediction issued at time $t$ can only explore data available up to time $t-\gamma$, where $\gamma$ is the delay associated with the availability of epidemiological data.

Assume that a given EEWS is required to issue predictions with time antecedence of $\beta$, i.e., at time $t$ the system would provide predictions associated with time $t + \beta$. Since it would not have access to epidemiological data associated with time between $t-\gamma$ and $t$, the EEWS may actually use data up to time $t - \gamma$ to forecast incidence at time $t+\beta$, effectively increasing the antecedence for which predictions are issued from $\beta$ to $\beta+\gamma$. We call this approach the \emph{increased antecedence approach}. The major disadvantage of this approach is that, the larger the antecedence, the more difficult it is to produce accurate forecasts, as incidence of infectious diseases during time $t$ tend to be strongly correlated with incidence in the recent past of $t$.

In this section, we describe an alternative to deal with delayed epidemiological data. We first provide few definitions key to understanding our proposal:
\begin{itemize}
  \item The current time is denoted by $t$, while epidemiological data delay is denoted by $\gamma$ and predictive antecedence by $\beta$.
  \item We call \emph{delayed epidemiological data} any epidemiological data associated with the time interval $t - \gamma$ and $t$.
  \item We call \emph{future epidemiological data} any epidemiological data associated with time $t' > t$.
\end{itemize}

The proposed approach for incorporating online data into EEWS is described in Figure \ref{fig:framework}. First, it uses a known relationship between online and epidemiological data (such as a linear or polynomial association) to estimate \emph{delayed epidemiological data} up to time $t$. Then, it evaluates how reliable are such estimates. If estimates are considered reliable, they are incorporated into available training set. Otherwise, they are simply ignored. Based on the potentially augmented training set, a model is estimated and used to forecast incidence at time $t + \beta$.

As indicated in Figure \ref{fig:framework}, the proposed framework is composed of three components. The first component is a predictive model that takes online data as a covariate (e.g., the number of dengue-related tweets) and epidemiological data as the response variable, and outputs estimates of delayed epidemiological data. The second component decides whether estimates of epidemiological data are safe to use, that is, whether the uncertainty associated to estimates is low. Uncertain estimates may introduce extra noise to the training set of the EEWS model, which can be detrimental. The simplest way to define whether estimated data is reliable is to use an uncertainty estimation threshold. Finally, the third component is a traditional EEWS model that takes as covariates data known to be associated with the disease (e.g., temperature, rainfall, time of the year) and epidemiological data as the response variable, and outputs future epidemiological data. 

\begin{figure*}[!t]
  \centering
  \subfigure{\includegraphics[width=0.99\linewidth]{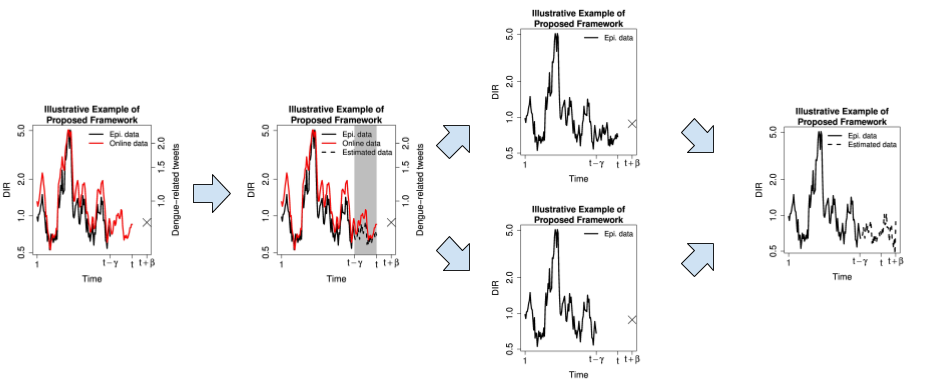}}
  \caption{An illustrative example of the use of the proposed framework. At the leftmost figure, we have actual DIR values up to time $t-\gamma$ and the number of dengue-related tweets up to time $t$ (solid black and red lines, respectively). Then, on the next figure, we estimate DIR values from time $t-\gamma$ to $t$ (dashed black line). Following, we have two possible scenarios: estimated DIR values are either incorporated into training data (top) or ignored (bottom). Finally, on the rightmost figure, estimates for future DIR values are made using the available training data.}
  \label{fig:example}
\end{figure*}

Figure~\ref{fig:example} illustrates the use of the framework. Given two time series, one corresponding to the epidemiological and the other to online data, note that the online data is available up to time $t$, while epidemiological data stops at $t -\gamma$. From these two series, we estimate the epidemiological data from $t -\gamma$ to $t$. Depending on the confidence of the estimated data, it can be incorporated to the training data of our spatial-temporal model or simply ignored.

The following sections describe how each component of the framework is defined to generate an accurate framework for dengue fever incidence forecasting.

\subsection{EEWS for Dengue Fever Incidence Forecasting}

We begin by defining the final component of the proposed framework: an epidemiological model to forecast future incidence based on the available training set, possibly extended by integrating estimated delayed DIR values.

Here, we define a model based on \citep{albinati2016accurate}, where the authors proposed to model log-transformed DIR values of Brazilian cities using a GP model equipped with a covariance function capable of exploiting temporal local dependences typical of infectious diseases, as well as the annual seasonality commonly observed in dengue data \citep{gharbi2011time,martinez2011predicting,promprou2006forecasting}. In summary, we propose to use the following two-part covariance function
\begin{eqnarray}
  k_s(t, t') & = & k_{loc}(t-t') + k_{qp}(t-t') \label{eq:cov} \\
  k_{loc}(\tau = t-t') & = & \sigma_{loc}^2 \left( 1 + \frac{\sqrt{5}\tau}{\ell_{loc}} + \frac{5\tau^2}{\ell_{loc}^2} \right) exp \left( - \frac{\sqrt{5}\tau}{\ell_{loc}} \right) \nonumber \\
  k_{qp}(\tau = t-t') & = & \sigma_{qp}^2 \left( 1 + \frac{\sqrt{5}\tau}{\ell_{qp}} + \frac{5\tau^2}{\ell_{qp}^2} \right) exp \left( - \frac{\sqrt{5}\tau}{\ell_{qp}} \right) \nonumber \\
  & & exp \left( -2 * \frac{sin^2(\pi |\tau| / p)}{\ell_{per}} \right) \nonumber
\end{eqnarray}
where $k_s(t,t)$ denotes the covariance between incidence values at city $s$ during weeks $t$ and $t'$ and $\sigma_{loc}$, $\ell_{loc}$, $\sigma_{qp}$, $\ell_{qp}$, $\ell_{per}$ and $p$ are hyperparameters learned from likelihood maximization.

\begin{figure}[h!]
  \centering
  \includegraphics[width=0.8\linewidth]{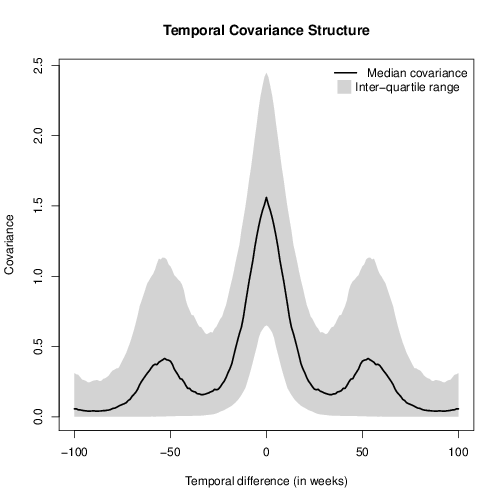}
  \caption{Covariance function considering temporal components in Equation \ref{eq:cov} with optimized hyperparameters according to Table \ref{tab:hypers}.}
  \label{fig:tempCorrelation}
\end{figure}

The intuition of the covariance function in Equation \ref{eq:cov} is the following. The function is formed by two temporal components, which are used to exploit two temporal patterns, namely local dependences and seasonality. Local dependences are exploited by $k_{loc}$, which correlates weeks nearby and is expressed in terms of a Mat\'ern covariance function, a covariance function commonly used in the GP literature to enforce smoothness by indicating that nearby data points are highly correlated, while distant data points are loosely correlated \citep{rasmussen2006gaussian}. Seasonality is exploited using a quasi-periodic function formulated as the product between a Mat\'ern and a periodic covariance function. By using this formulation, the model is capable of using information from incidence of past years, but is forced to give more relevance to more recent years than very distant ones. Figure \ref{fig:tempCorrelation} shows the covariance function obtained after hyperparameter optimization (Table \ref{tab:hypers}). Note that covariance is high for $\tau$ close to 0, with smaller peaks for $\tau$ around 1 year (52 weeks).

\begin{table}[h!]
  \centering
  \caption{Hyperparameters obtained after likelihood maximization}
  \label{tab:hypers}
  \begin{tabular}{cccc}
    \hline
    & 1st Quartile & Median & 3rd Quartile \\
    \hline
    $\sigma_{loc}^2$ & 0.039 & 0.101 & 0.186 \\
    $\sigma_{qp}^2$ & 0.484 & 1.427 & 2.311 \\
    $\ell_{loc}$ & 1.000 & 2.000 & 7.000 \\
    $\ell_{qp}$ & 41.000 & 58.000 & 99.000 \\
    $\ell_{per}$ & 1.000 & 1.000 & 2.000 \\
    $p$ & 54.000 & 59.000 & 73.000 \\
    \hline
  \end{tabular}
\end{table}

Having defined the covariance function, the model used in this work can be expressed as follows:
\begin{eqnarray}
  DIR_{s,t} & = & exp\left( y_{s,t} + \overline{y_s} \right) - 1 \label{eq:eews} \\
  y_{s,\cdot} & \sim & \mathcal{GP} \left( 0, k_s(t,t') \right) \nonumber
\end{eqnarray}
where $DIR_{s,t}$ denotes the DIR value at city $s$ during week $t$, $\overline{y_s}$ indicates the mean of log-transformed DIR values at city $s$ and $k_s(t,t')$ is the covariance function indicated in Equation \ref{eq:cov}.

\subsection{Estimating Delayed Epidemiological Data}

The first step on the proposed framework is to exploit a known relationship between online and epidemiological data. Based on the study provided by \citet{gomide2011dengue}, which identified a linear association between the number of dengue-related tweets and DIR, and on the results obtained by the Gaussian process-based model proposed by \citet{albinati2016accurate}, we opted for a Gaussian process model equipped with the covariance function indicated in Equation \ref{eq:cov}, with an additional component designed to exploit the linear association between Twitter and epidemiological data.

Following \citet{albinati2016accurate}, we first applied a logarithmic transformation of DIR values to avoid using Gaussian distributions to model count data. We then centered the resulting values and estimated the zero-mean Gaussian process model. In summary, this component is implemented by the following model:
\begin{eqnarray}
  DIR_{s,t} & = & exp(y_{s,t} + \overline{y_s}) - 1 \label{eq:comp1} \\
  y_{s,\cdot} & \sim & \mathcal{GP} \left( 0, k_s(t,t') + \frac{x_{s,t} * x_{s,t'}}{\ell_{tw}} \right) \nonumber
\end{eqnarray}
where $x_{s,t}$ indicates the log-transformed number of dengue-related tweets posted during week $t$ at city $s$ and $k_s(t,t')$ is given by Equation \ref{eq:cov}. Note that $\ell_{tw}$ is an additional hyperparameter, which is also learned from likelihood maximization using available training data.

\subsection{Deciding Whether to Use Estimates}

In order to avoid using poorly estimated DIR values, which could introduce noise into epidemiological data, we defined an extra module responsible for deciding whether estimated epidemiological data is reliable or not. Since we are assuming a linear relationship between Twitter and dengue data, we opted for using a threshold based on the correlation between epidemiological and Twitter data. Whenever the correlation exceeds the threshold, estimated dengue data is considered. Otherwise, only actual epidemiological dengue data is used. Details on how to define this threshold are given on Section \ref{sec:experiments}.


%% file: experiments.tex
In this section, we show empirical results obtained by the proposed framework. We first define our general experimental setup, and then assess the accuracy of the proposed framework for incorporating Twitter data into dengue EEWSs by verifying if it leads to more accurate predictions than the increased antecedence approach. Then, we provide some discussion on the predictions obtained and on the appropriateness of the proposed framework.

\subsection{Experimental Setup}

Motivated by the incidence levels defined by the Brazilian Ministry of Health, we evaluate each model according to the area under the receiver operating characteristic curve (AUC) considering incidence levels as labels. The receiver operating characteristic curve is created by plotting the true positive rate against the false positive rate at multiple threshold values, used for defining whether the predicted class is positive or negative. In this sense, it indicates the trade-off between sensitivity and specificity. An area close to 1 indicates that one can increase the model's sensitivity by changing the threshold incurring in few false positive errors, while smaller areas indicate more false positive errors when increasing sensitivity. For each week and city under study, we verify the attained and predicted incidence levels. We calculate the AUC for each level on an one-against-all strategy and then compute the average over the three levels.

Predictions were issued always with $\beta = 4$ weeks of antecedence. Choosing the antecedence value implies in dealing with a trade-off: shorter antecedence leads to more data and potentially more accurate predictions, but less time for health authorities to act in advance to mitigate the impact of future outbreaks. We chose $\beta = 4$ weeks as a balance between utility and accuracy, allowing some time for health authorities to act while still managing to obtain accurate predictions.

We also defined that the first two years (weeks 1 to 104) would be used for training only. Predictions are issued from the third to the sixth year, always with 4 weeks in advance, as indicated previously. This means that our available training set is continuously growing with time, as we collect more data.

Finally, models are always compared using a paired Wilcoxon test with 95\% confidence level. Bonferroni corrections were applied whenever multiple statistical tests were required to ensure statistical significance.

\subsection{Evaluation of Proposed Framework}

\begin{figure}[h!]
  \centering
  \includegraphics[width=0.7\linewidth]{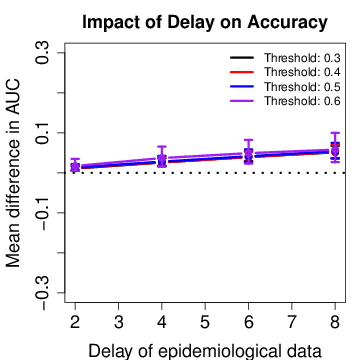}
  \caption{Mean difference in accuracy when comparing the proposed framework with extending predictive antecedence to deal with epidemiological data delay. Whiskers indicate 95\% confidence intervals.}
  \label{fig:difference}
\end{figure}

In order to use the proposed framework, we must define the correlation threshold that defines whether estimated DIR values are going to be integrated into the training set or not. For that, we instantiated four versions of the proposed framework, each with a distinct threshold value. We compared the difference in accuracy assuming four values of delay in the epidemiological data for each threshold used when compared to simply extending predictive antecedence, as shown in Figure \ref{fig:difference}. Positive values indicate that the proposed framework achieved better results, while negative values indicate the opposite scenario. Note that, although it is difficult to identify the best threshold value, the proposed framework obtained more accurate predictions for all delays and threshold values used in the experiment, obtaining statistically superior results for delays of at least 4 weeks. Besides that, the difference seems to grow with larger delays, indicating that the proposed framework is indeed more robust to delays in epidemiological data.

\begin{figure}[h!]
  \centering
  \includegraphics[width=0.7\linewidth]{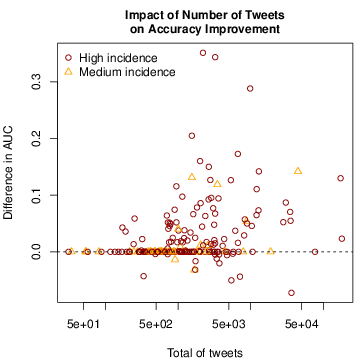}
  \caption{Impact of the amount of Twitter data in the observed difference of accuracy when considering predictions provided by the proposed methodology and provided by the original model with extended predictive antecedence. Each symbol denotes a city, with its shape and color indicating the highest incidence level it achieved in the period under study.}
  \label{fig:differenceNumberTweets}
\end{figure}

Having verified that the proposed framework leads to more accurate predictions in general, we would like to better understand which cities are improved by the proposed methodology and which cities are not. For this analysis, we considered the threshold value of 0.5 and fixed the epidemiological delay as 6 weeks to investigate the role of the amount of Twitter data available on the accuracy improvement obtained. Figure \ref{fig:differenceNumberTweets} shows that cities that obtained higher improvements in accuracy tend to have more Twitter data available. This result corroborates with the intuitive notion that more Twitter data eases the identification of stronger relationships with epidemiological dengue data.

\begin{figure}[h!]
  \centering
  \includegraphics[width=0.9\linewidth]{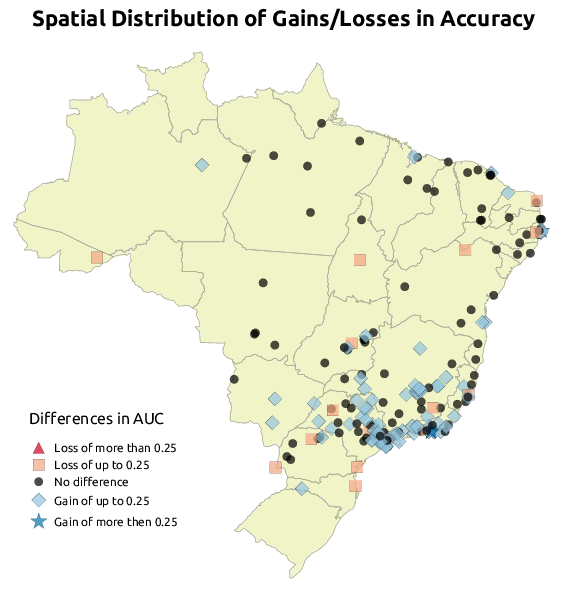}
  \caption{Spatial distribution of the observed difference of accuracy when considering predictions provided by the proposed methodology and provided by the original model with extended predictive antecedence. Each symbol denotes a city, with its shape and color indicating the gain/loss in accuracy provided by the proposed methodology.}
  \label{fig:differenceSpatialDist}
\end{figure}

We now evaluate the spatial distribution of accuracy improvements. Figure \ref{fig:differenceSpatialDist} shows the spatial distribution of accuracy improvements obtained when applying the proposed framework. Note that improvements are mostly concentrated within the Southeast region of Brazil. Since the Southeast region of Brazil is the most developed region of the country, this distribution reflects the availability of Internet access in Brazil. Having more people capable of posting on Twitter, this region was the most benefited by the proposed framework. We also highlight that, for only 17 cities out of 213, applying the proposed framework was detrimental, indicating that the proposed framework is safe to use regardless of where cities are located.

\begin{figure}[h!]
  \centering
  \includegraphics[width=0.7\linewidth]{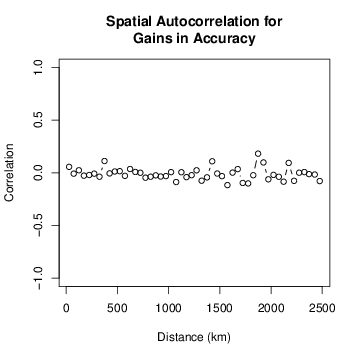}
  \caption{Spatial autocorrelation of the observed difference of accuracy when considering predictions provided by the proposed methodology and provided by the original model with extended predictive antecedence.}
  \label{fig:differenceSpatialCor}
\end{figure}

Although bigger improvements in accuracy where concentrated in the Southeast region of Brazil, we did not observe significant spatial autocorrelation between improvements, as shown in Figure \ref{fig:differenceSpatialCor}.

\subsection{Analysis of Predictions}

In the last section, we evaluated whether the proposed methodology led to more accurate predictions than the increased antecedence approach. We now evaluate the appropriateness of the proposed methodology and analyze the final predictions issued by it.

\begin{figure}[h!]
  \centering
  \includegraphics[width=0.7\linewidth]{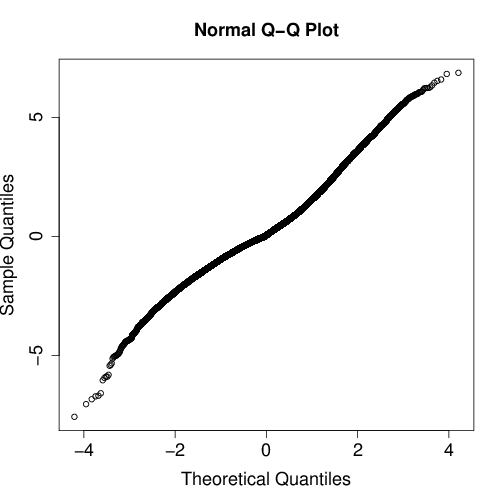}
  \caption{Quantile-quantile normal plot of residuals.}
  \label{fig:qqplot}
\end{figure}

We begin by evaluating two assumptions of the proposed model: homoscedasticity and normality of residuals. In other words, the GP-based model assumes that residuals are Gaussian, independent and identically distributed. Figure \ref{fig:qqplot} shows the quantile-quantile plot of residuals. Note that observed quantiles are associated with theoretical quantiles in a approximately linear fashion, indicating that residuals are approximately Gaussian, independent and identically distributed.

\begin{figure}[h!]
  \centering
  \includegraphics[width=0.7\linewidth]{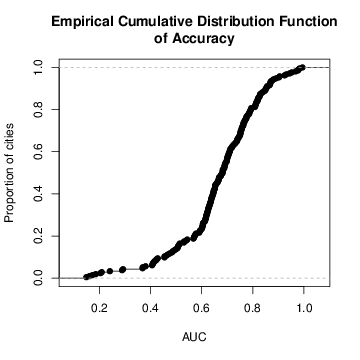}
  \caption{Empirical cumulative distribution function of the observed accuracy when considering predictions provided by the proposed methodology.}
  \label{fig:cdfAUC}
\end{figure}

Figure \ref{fig:cdfAUC} shows empirical cumulative distribution function of AUC. Almost 77\% of the cities obtained AUC higher than 0.6, while about 20\% of the cities obtained AUC higher than 0.8.

\begin{figure}[h!]
  \centering
  \includegraphics[width=0.7\linewidth]{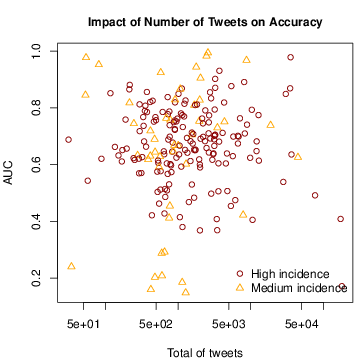}
  \caption{Impact of the amount of Twitter data in the observed accuracy when considering predictions provided by the proposed methodology.}
  \label{fig:aucNumberTweets}
\end{figure}

Figure \ref{fig:aucNumberTweets}, on the other hand, shows the impact of the volume of Twitter data on the accuracy of predictions. Note that the amount of Twitter data is not associated with final predictions. We believe this result is mainly due to the module deciding whether to use estimated DIR values or not, which makes the whole methodology safe to use even when Twitter data is scarce.

\begin{figure}[h!]
  \centering
  \includegraphics[width=0.9\linewidth]{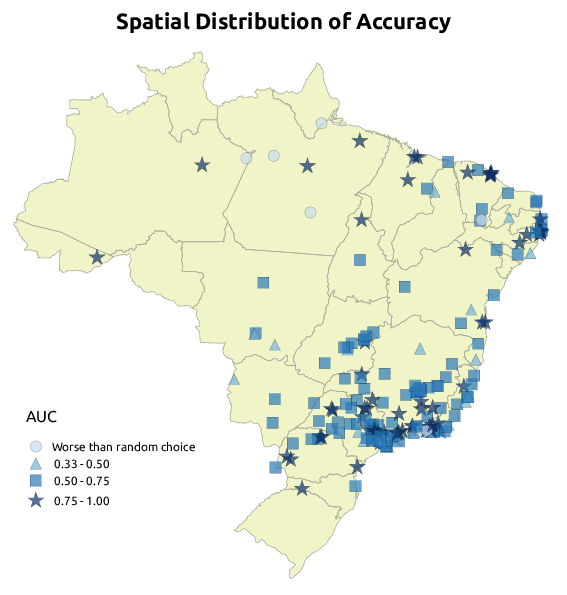}
  \caption{Spatial distribution of the observed accuracy when considering predictions provided by the proposed methodology.}
  \label{fig:aucSpatialDist}
\end{figure}

The spatial distribution of final accuracies is shown in Figure \ref{fig:aucSpatialDist}. Differently from Figure \ref{fig:differenceSpatialDist}, high final accuracies are not strongly concentrated only within the Southeast region. In fact, a significant number of cities from North and Northeast regions obtained high AUC values. These results suggest that, although Twitter was not strongly helpful for cities within these regions, the GP-based EEWS is still capable of issuing accurate predictions with increased predictive antecedence. In this direction, one could argue that the proposed framework is capable of improving predictions exactly where delays in epidemiological data are more detrimental to EEWSs.

\begin{figure*}
  \centering
  \includegraphics[width=0.9\linewidth]{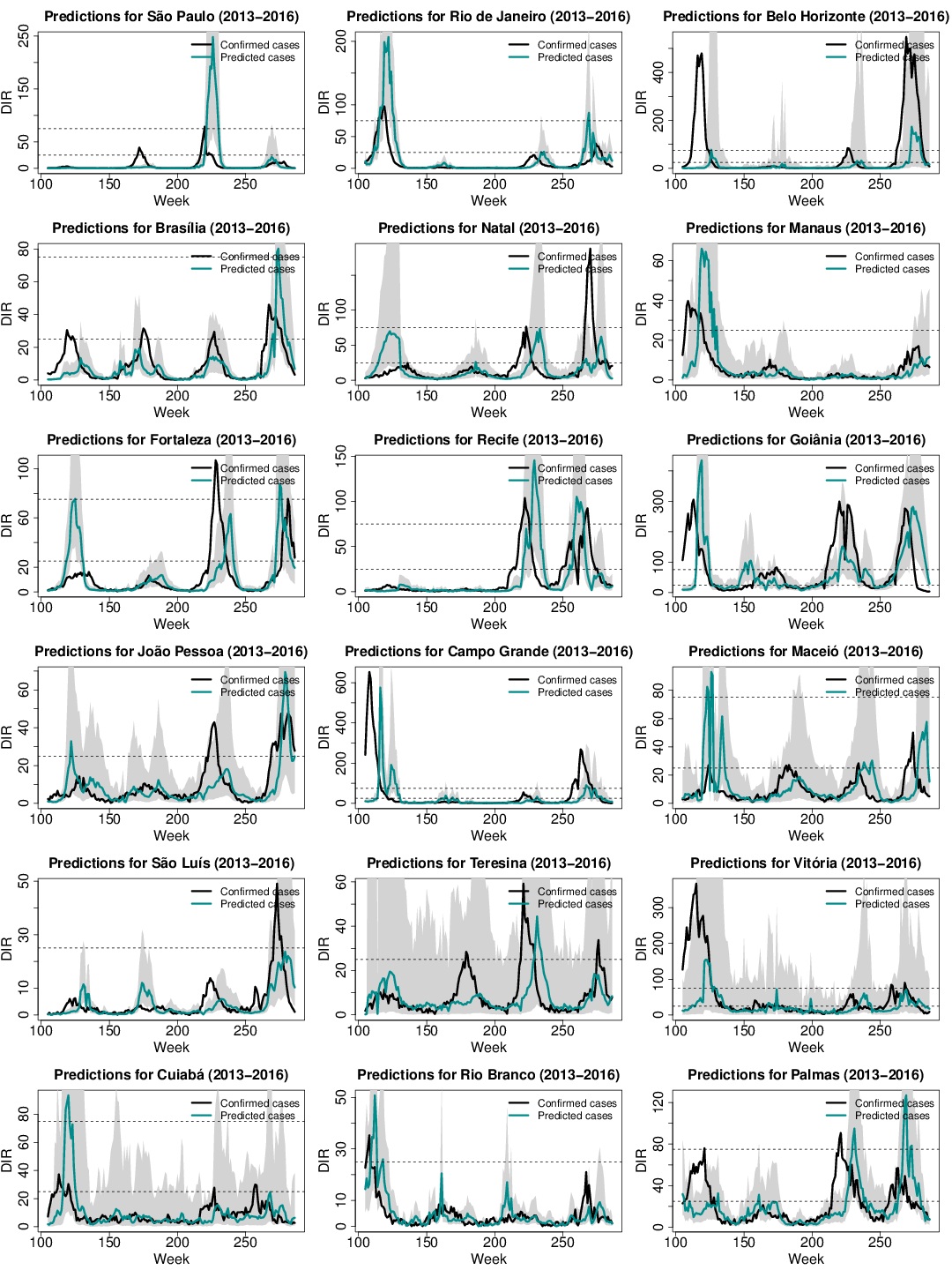}
  \caption{Predictions issued for all Brazilian capital cities under study. The solid black line denotes the observed DIR, while the cyan line indicates the predicted DIR. The gray shaded area exhibits the 95\% confidence interval. Dashed lines indicate the boundaries between incidence levels.}
  \label{fig:predictions}
\end{figure*}

Figure \ref{fig:predictions} shows predictions issued by the proposed framework for all Brazilian capital cities that reached at least medium incidence from 2013 to 2016. Cities are sorted by the number of dengue-related tweets collected, where S\~{a}o Paulo was the city the with highest number of tweets and Palmas was the city with the lowest number of tweets. Note that, even for cities with low amount of Twitter data, the proposed framework typically detects major dengue fever outbreaks, reaching the correct incidence level (indicated by the dashed lines). The city of Belo Horizonte, although being the capital city with third highest number of dengue-related tweets, demonstrated to be particularly challenging to forecast, since it exhibited very low incidence during all period under study, except from the beginning of 2013 and 2016, where very drastic outbreaks were observed. However, we note that our proposed methodology, although failing to capture the outbreak in 2013, was able to capture the increase in incidence associated with the outbreak in 2016, reaching the correct incidence level.

%% file: conclusions.tex
Although EEWS can be employed to forecast DIR, they require up-to-date epidemiological data to work effectively. Unfortunately, epidemiological data requires time to be available, since it may require time-consuming laboratorial tests. Therefore, at any given moment in time $t$, we cannot expect to have epidemiological data available up to current time. Instead, we would expect to have data available up to time $t-\gamma$. This lack of up-to-date epidemiological data can be highly detrimental to EEWS, as data associated to a recent past is typically very informative, specially for infectious diseases.

In order to mitigate this issue, this work proposes a framework that exploits known relationships between epidemiological and online data, such as data coming from Twitter or Facebook, to estimate current epidemiological data. Effective EEWS can then be built based on these estimates, without the need of increasing predictive antecedence.

We show that the proposed framework leads to more accurate predictions than employing an EEWS with larger predictive antecedence to account for the delay of epidemiological data. In fact, for delays greater than or equal to 4 weeks, the proposed framework provided statistically more accurate predictions in general. We also show that, although larger amounts of online data led to larger improvements in accuracy, the framework is robust in the sense of being capable of identifying most of the cities for which it was not worthwhile to use Twitter data. For these cities, estimating delayed epidemiological data would introduce noise to the training dataset, thus being detrimental to accuracy. In the end, the proposed framework was able to obtain accurate predictions, with majority of cities obtaining an area under the ROC curve higher than 0.68, when considering that epidemiological data is delayed by 6 weeks.

Our proposed framework obtained higher improvements in accuracy for cities within the Southeast region. According to a census conducted by the Brazilian state-run agency IBGE\footnote{Brazilian Institute of Geography and Statistics (Instituto Brasileiro de Geografia e Estatística, in Portuguese) -- http://www.ibge.gov.br}, in 2014, the Southeast region was the region of more houses with Internet access (62.9\%), while about only 55\% of the Brazilian homes have access to Internet \citep{ibge2014pnad}. With more access to the Internet, people within this region are able to produce more dengue-related tweets, which we show to be related to the improvement obtained by the proposed framework. Our experiments, however, suggested that this region is the most affected by the lack of up-to-date epidemiological data. Besides that, it concentrates the majority of dengue cases in Brazil, therefore being one of the Brazilian regions where the proposed framework would be most helpful.

Although improving accuracy in general, the proposed framework has its limitations. First, some cities were not improved due to a weak relationship between online data and epidemiological data. We believe this fact is due to two major reasons: (i) data scarcity and (ii) non-linear associations between dengue and Twitter data. To deal with the former issue, as future work, we intend to aggregate other online data sources, such as data from Google Trends and Wikipedia access logs, which are already being used to estimate current epidemiological data of some diseases. For the latter, we aim to evaluate other models to estimate delayed epidemiological data.

It is important to stress that, although we obtained good predictive accuracy for a large set of Brazilian cities, decision making by public health authorities should not be solely based on estimates provided by the proposed framework, which can be affected by fluctuations on Twitter, among other effects. In fact, our main objective was to develop a system capable of identifying potential future outbreaks, focusing the attention of specialists to specific regions of the country. By doing so, we hope to facilitate and improve their work, since they would be able to better investigate a smaller subset of areas, instead of tracking down DIR around the whole country.

Finally, we would like to highlight that we believe that the proposed framework, although evaluated in the scenario of dengue fever in Brazilian cities, is flexible and may be applied to other scenarios without large modifications. For addressing this, as future work, we intend to apply it to other diseases, specially mosquito-borne diseases, which typically exhibit similar temporal patterns to dengue fever.

%% file: main.bbl

\begin{thebibliography}{00}


\ifx \showCODEN    \undefined \def \showCODEN     #1{\unskip}     \fi
\ifx \showDOI      \undefined \def \showDOI       #1{#1}\fi
\ifx \showISBNx    \undefined \def \showISBNx     #1{\unskip}     \fi
\ifx \showISBNxiii \undefined \def \showISBNxiii  #1{\unskip}     \fi
\ifx \showISSN     \undefined \def \showISSN      #1{\unskip}     \fi
\ifx \showLCCN     \undefined \def \showLCCN      #1{\unskip}     \fi
\ifx \shownote     \undefined \def \shownote      #1{#1}          \fi
\ifx \showarticletitle \undefined \def \showarticletitle #1{#1}   \fi
\ifx \showURL      \undefined \def \showURL       {\relax}        \fi
\providecommand\bibfield[2]{#2}
\providecommand\bibinfo[2]{#2}
\providecommand\natexlab[1]{#1}
\providecommand\showeprint[2][]{arXiv:#2}

\bibitem[\protect\citeauthoryear{Albinati, Meira~Jr, and Pappa}{Albinati
  et~al\mbox{.}}{2016}]%
        {albinati2016accurate}
\bibfield{author}{\bibinfo{person}{Julio Albinati}, \bibinfo{person}{Wagner
  Meira~Jr}, {and} \bibinfo{person}{Gisele~Lobo Pappa}.}
  \bibinfo{year}{2016}\natexlab{}.
\newblock \showarticletitle{An Accurate Gaussian Process-Based Early Warning
  System for Dengue Fever}. In \bibinfo{booktitle}{{\em 2016 Brazilian
  Conference on Intelligent Systems, {BRACIS} 2016, Recife, Brazil, October
  10-13, 2016}}.
\newblock


\bibitem[\protect\citeauthoryear{Althouse, Ng, and Cummings}{Althouse
  et~al\mbox{.}}{2011}]%
        {althouse2011prediction}
\bibfield{author}{\bibinfo{person}{Benjamin~M. Althouse},
  \bibinfo{person}{Yih~Yng Ng}, {and} \bibinfo{person}{Derek A.~T. Cummings}.}
  \bibinfo{year}{2011}\natexlab{}.
\newblock \showarticletitle{Prediction of Dengue Incidence Using Search Query
  Surveillance}.
\newblock \bibinfo{journal}{{\em PLoS Negl Trop Dis\/}} \bibinfo{volume}{5},
  \bibinfo{number}{8} (\bibinfo{date}{08} \bibinfo{year}{2011}),
  \bibinfo{pages}{1--7}.
\newblock
\showDOI{%
\url{https://doi.org/10.1371/journal.pntd.0001258}}


\bibitem[\protect\citeauthoryear{Broniatowski, Paul, and Dredze}{Broniatowski
  et~al\mbox{.}}{2013}]%
        {broniatowski2013national}
\bibfield{author}{\bibinfo{person}{David~A. Broniatowski},
  \bibinfo{person}{Michael~J. Paul}, {and} \bibinfo{person}{Mark Dredze}.}
  \bibinfo{year}{2013}\natexlab{}.
\newblock \showarticletitle{National and Local Influenza Surveillance through
  Twitter: An Analysis of the 2012-2013 Influenza Epidemic}.
\newblock \bibinfo{journal}{{\em PLoS ONE\/}} \bibinfo{volume}{8},
  \bibinfo{number}{12} (\bibinfo{date}{12} \bibinfo{year}{2013}).
\newblock
\showDOI{%
\url{https://doi.org/10.1371/journal.pone.0083672}}


\bibitem[\protect\citeauthoryear{Dom, Hassan, Latif, and Ismail}{Dom
  et~al\mbox{.}}{2013}]%
        {dom2013generating}
\bibfield{author}{\bibinfo{person}{Nazri~Che Dom}, \bibinfo{person}{A~Abu
  Hassan}, \bibinfo{person}{Z~Abd Latif}, {and} \bibinfo{person}{Rodziah
  Ismail}.} \bibinfo{year}{2013}\natexlab{}.
\newblock \showarticletitle{Generating temporal model using climate variables
  for the prediction of dengue cases in Subang Jaya, Malaysia}.
\newblock \bibinfo{journal}{{\em Asian Pacific Journal of Tropical Disease\/}}
  \bibinfo{volume}{3}, \bibinfo{number}{5} (\bibinfo{year}{2013}),
  \bibinfo{pages}{352 -- 361}.
\newblock
\showISSN{2222-1808}
\showDOI{%
\url{https://doi.org/10.1016/S2222-1808(13)60084-5}}


\bibitem[\protect\citeauthoryear{Eastin, Delmelle, Casas, Wexler, and
  Self}{Eastin et~al\mbox{.}}{2014}]%
        {eastin2014intra}
\bibfield{author}{\bibinfo{person}{Matthew~D Eastin}, \bibinfo{person}{Eric
  Delmelle}, \bibinfo{person}{Irene Casas}, \bibinfo{person}{Joshua Wexler},
  {and} \bibinfo{person}{Cameron Self}.} \bibinfo{year}{2014}\natexlab{}.
\newblock \showarticletitle{Intra-and interseasonal autoregressive prediction
  of dengue outbreaks using local weather and regional climate for a tropical
  environment in Colombia}.
\newblock \bibinfo{journal}{{\em The American journal of tropical medicine and
  hygiene\/}} \bibinfo{volume}{91}, \bibinfo{number}{3} (\bibinfo{year}{2014}),
  \bibinfo{pages}{598--610}.
\newblock


\bibitem[\protect\citeauthoryear{Generous, Fairchild, Deshpande, Del~Valle, and
  Priedhorsky}{Generous et~al\mbox{.}}{2014}]%
        {generous2014global}
\bibfield{author}{\bibinfo{person}{Nicholas Generous},
  \bibinfo{person}{Geoffrey Fairchild}, \bibinfo{person}{Alina Deshpande},
  \bibinfo{person}{Sara~Y. Del~Valle}, {and} \bibinfo{person}{Reid
  Priedhorsky}.} \bibinfo{year}{2014}\natexlab{}.
\newblock \showarticletitle{Global Disease Monitoring and Forecasting with
  Wikipedia}.
\newblock \bibinfo{journal}{{\em PLoS Comput Biol\/}} \bibinfo{volume}{10},
  \bibinfo{number}{11} (\bibinfo{date}{11} \bibinfo{year}{2014}),
  \bibinfo{pages}{1--16}.
\newblock
\showDOI{%
\url{https://doi.org/10.1371/journal.pcbi.1003892}}


\bibitem[\protect\citeauthoryear{Gharbi, Quenel, Gustave, Cassadou, La~Ruche,
  Girdary, and Marrama}{Gharbi et~al\mbox{.}}{2011}]%
        {gharbi2011time}
\bibfield{author}{\bibinfo{person}{Myriam Gharbi}, \bibinfo{person}{Philippe
  Quenel}, \bibinfo{person}{Jo{\"e}l Gustave}, \bibinfo{person}{Sylvie
  Cassadou}, \bibinfo{person}{Guy La~Ruche}, \bibinfo{person}{Laurent Girdary},
  {and} \bibinfo{person}{Laurence Marrama}.} \bibinfo{year}{2011}\natexlab{}.
\newblock \showarticletitle{Time series analysis of dengue incidence in
  Guadeloupe, French West Indies: forecasting models using climate variables as
  predictors}.
\newblock \bibinfo{journal}{{\em BMC infectious diseases\/}}
  \bibinfo{volume}{11} (\bibinfo{year}{2011}).
\newblock


\bibitem[\protect\citeauthoryear{Gomide, Veloso, Meira, Almeida, Benevenuto,
  Ferraz, and Teixeira}{Gomide et~al\mbox{.}}{2011}]%
        {gomide2011dengue}
\bibfield{author}{\bibinfo{person}{Jana\'{\i}na Gomide},
  \bibinfo{person}{Adriano Veloso}, \bibinfo{person}{Wagner Meira, Jr.},
  \bibinfo{person}{Virg\'{\i}lio Almeida}, \bibinfo{person}{Fabr\'{\i}cio
  Benevenuto}, \bibinfo{person}{Fernanda Ferraz}, {and} \bibinfo{person}{Mauro
  Teixeira}.} \bibinfo{year}{2011}\natexlab{}.
\newblock \showarticletitle{Dengue Surveillance Based on a Computational Model
  of Spatio-temporal Locality of Twitter}. In \bibinfo{booktitle}{{\em
  Proceedings of the 3rd International Web Science Conference}} {\em
  (\bibinfo{series}{WebSci '11})}. \bibinfo{publisher}{ACM},
  \bibinfo{address}{New York, NY, USA}, Article \bibinfo{articleno}{3},
  \bibinfo{numpages}{8}~pages.
\newblock
\showISBNx{978-1-4503-0855-7}
\showDOI{%
\url{https://doi.org/10.1145/2527031.2527049}}


\bibitem[\protect\citeauthoryear{{Instituto Brasileiro de Geografia e
  Estatística}}{{Instituto Brasileiro de Geografia e Estatística}}{2014}]%
        {ibge2014pnad}
\bibfield{author}{\bibinfo{person}{{Instituto Brasileiro de Geografia e
  Estatística}}.} \bibinfo{year}{2014}\natexlab{}.
\newblock \bibinfo{title}{Pesquisa Nacional por Amostra de Domicílios -- PNAD
  2014}.
\newblock
  \bibinfo{howpublished}{http://ibge.gov.br/home/estatistica/populacao/acessoainternet2014/default\_xls.shtm}.
    (\bibinfo{year}{2014}).
\newblock
\newblock
\shownote{[Online; accessed 18-April-2017; in Portuguese].}


\bibitem[\protect\citeauthoryear{Martinez and Silva}{Martinez and
  Silva}{2011}]%
        {martinez2011predicting}
\bibfield{author}{\bibinfo{person}{Edson~Zangiacomi Martinez} {and}
  \bibinfo{person}{Elis{\^a}ngela Aparecida Soares~da Silva}.}
  \bibinfo{year}{2011}\natexlab{}.
\newblock \showarticletitle{Predicting the number of cases of dengue infection
  in Ribeir{\~a}o Preto, S{\~a}o Paulo State, Brazil, using a SARIMA model}.
\newblock \bibinfo{journal}{{\em Cadernos de saude publica\/}}
  \bibinfo{volume}{27}, \bibinfo{number}{9} (\bibinfo{year}{2011}),
  \bibinfo{pages}{1809--1818}.
\newblock


\bibitem[\protect\citeauthoryear{Promprou, Jaroensutasinee, and
  Jaroensutasinee}{Promprou et~al\mbox{.}}{2006}]%
        {promprou2006forecasting}
\bibfield{author}{\bibinfo{person}{S Promprou}, \bibinfo{person}{M
  Jaroensutasinee}, {and} \bibinfo{person}{K Jaroensutasinee}.}
  \bibinfo{year}{2006}\natexlab{}.
\newblock \showarticletitle{Forecasting dengue haemorrhagic fever cases in
  Southern Thailand using ARIMA Models}.
\newblock \bibinfo{journal}{{\em Dengue Bulletin\/}}  \bibinfo{volume}{30}
  (\bibinfo{year}{2006}), \bibinfo{pages}{99}.
\newblock


\bibitem[\protect\citeauthoryear{Rasmussen and Williams}{Rasmussen and
  Williams}{2006}]%
        {rasmussen2006gaussian}
\bibfield{author}{\bibinfo{person}{Carl~Edward Rasmussen} {and}
  \bibinfo{person}{Christopher K.~I. Williams}.}
  \bibinfo{year}{2006}\natexlab{}.
\newblock \bibinfo{booktitle}{{\em Gaussian Processes for Machine Learning}}.
\newblock \bibinfo{publisher}{The MIT Press}, \bibinfo{address}{Massachusetts
  Institute of Technology, Cambridge, Massachusetts 02142}.
\newblock
\showISBNx{0-262-18253-X}


\bibitem[\protect\citeauthoryear{Sakaki, Okazaki, and Matsuo}{Sakaki
  et~al\mbox{.}}{2010}]%
        {sakaki2010eathquake}
\bibfield{author}{\bibinfo{person}{Takeshi Sakaki}, \bibinfo{person}{Makoto
  Okazaki}, {and} \bibinfo{person}{Yutaka Matsuo}.}
  \bibinfo{year}{2010}\natexlab{}.
\newblock \showarticletitle{Earthquake Shakes Twitter Users: Real-time Event
  Detection by Social Sensors}. In \bibinfo{booktitle}{{\em Proceedings of the
  19th International Conference on World Wide Web}} {\em (\bibinfo{series}{WWW
  '10})}. \bibinfo{publisher}{ACM}, \bibinfo{address}{New York, NY, USA},
  \bibinfo{pages}{851--860}.
\newblock
\showISBNx{978-1-60558-799-8}
\showDOI{%
\url{https://doi.org/10.1145/1772690.1772777}}


\bibitem[\protect\citeauthoryear{Samir, Gething, Brady, Messina, Farlow, Moyes,
  Drake, Brownstein, Hoen, Sankoh, Myers, George, Jaenisch, Wint, Simmons,
  Scott, Farrar, and Hay}{Samir et~al\mbox{.}}{2013}]%
        {samir2013global}
\bibfield{author}{\bibinfo{person}{Bhatt Samir}, \bibinfo{person}{Peter~W.
  Gething}, \bibinfo{person}{Oliver~J. Brady}, \bibinfo{person}{Jane~P.
  Messina}, \bibinfo{person}{Andrew~W. Farlow}, \bibinfo{person}{Catherine~L.
  Moyes}, \bibinfo{person}{John~M. Drake}, \bibinfo{person}{John~S.
  Brownstein}, \bibinfo{person}{Anne~G. Hoen}, \bibinfo{person}{Osman Sankoh},
  \bibinfo{person}{Monica~F. Myers}, \bibinfo{person}{Dylan~B. George},
  \bibinfo{person}{Thomas Jaenisch}, \bibinfo{person}{G.~R.~William Wint},
  \bibinfo{person}{Cameron~P. Simmons}, \bibinfo{person}{Thomas~W. Scott},
  \bibinfo{person}{Jeremy~J. Farrar}, {and} \bibinfo{person}{Simon~I. Hay}.}
  \bibinfo{year}{2013}\natexlab{}.
\newblock \showarticletitle{{The global distribution and burden of dengue}}.
\newblock \bibinfo{journal}{{\em Nature\/}} \bibinfo{volume}{496},
  \bibinfo{number}{7446} (\bibinfo{date}{apr} \bibinfo{year}{2013}),
  \bibinfo{pages}{504–507}.
\newblock
\showISSN{0028-0836}
\showURL{%
\url{http://www.nature.com/nature/journal/v496/n7446/abs/nature12060.html\#supplementary-information}}


\bibitem[\protect\citeauthoryear{Shaman, Karspeck, Yang, Tamerius, and
  Lipsitch}{Shaman et~al\mbox{.}}{2013}]%
        {shaman2013real}
\bibfield{author}{\bibinfo{person}{Jeffrey Shaman}, \bibinfo{person}{Alicia
  Karspeck}, \bibinfo{person}{Wan Yang}, \bibinfo{person}{James Tamerius},
  {and} \bibinfo{person}{Marc Lipsitch}.} \bibinfo{year}{2013}\natexlab{}.
\newblock \showarticletitle{Real-time influenza forecasts during the 2012--2013
  season}.
\newblock \bibinfo{journal}{{\em Nature communications\/}}  \bibinfo{volume}{4}
  (\bibinfo{year}{2013}).
\newblock


\bibitem[\protect\citeauthoryear{Souza, de~Brito, Assun{\c{c}}{\~a}o, and
  Meira~Jr}{Souza et~al\mbox{.}}{2015}]%
        {souza2015latent}
\bibfield{author}{\bibinfo{person}{Roberto~CSNP Souza},
  \bibinfo{person}{Denise~EF de Brito}, \bibinfo{person}{Renato~M
  Assun{\c{c}}{\~a}o}, {and} \bibinfo{person}{Wagner Meira~Jr}.}
  \bibinfo{year}{2015}\natexlab{}.
\newblock \showarticletitle{A latent shared-component generative model for
  real-time disease surveillance using Twitter data}.
\newblock \bibinfo{journal}{{\em arXiv preprint arXiv:1510.05981\/}}
  (\bibinfo{year}{2015}).
\newblock


\bibitem[\protect\citeauthoryear{Souza, de~Brito, Cardoso, de~Oliveira, Jr.,
  and Pappa}{Souza et~al\mbox{.}}{2014}]%
        {souza2014evolutionary}
\bibfield{author}{\bibinfo{person}{Roberto C. S. N.~P. Souza},
  \bibinfo{person}{Denise E.~F. de Brito}, \bibinfo{person}{Rodrigo~L.
  Cardoso}, \bibinfo{person}{Derick~M. de Oliveira},
  \bibinfo{person}{Wagner~Meira Jr.}, {and} \bibinfo{person}{Gisele~L. Pappa}.}
  \bibinfo{year}{2014}\natexlab{}.
\newblock \showarticletitle{An Evolutionary Methodology for Handling Data
  Scarcity and Noise in Monitoring Real Events from Social Media Data}. In
  \bibinfo{booktitle}{{\em Advances in Artificial Intelligence - {IBERAMIA}
  2014 - 14th Ibero-American Conference on AI, Santiago de Chile, Chile,
  November 24-27, 2014, Proceedings}} {\em (\bibinfo{series}{Lecture Notes in
  Computer Science})}, \bibfield{editor}{\bibinfo{person}{Ana L.~C. Bazzan}
  {and} \bibinfo{person}{Karim Pichara}} (Eds.), Vol.~\bibinfo{volume}{8864}.
  \bibinfo{publisher}{Springer}, \bibinfo{pages}{295--306}.
\newblock
\showDOI{%
\url{https://doi.org/10.1007/978-3-319-12027-0_24}}


\bibitem[\protect\citeauthoryear{Wiltshire}{Wiltshire}{2006}]%
        {wiltshire2006developing}
\bibfield{author}{\bibinfo{person}{Alison Wiltshire}.}
  \bibinfo{year}{2006}\natexlab{}.
\newblock \showarticletitle{‘Developing early warning systems: A checklist}.
  In \bibinfo{booktitle}{{\em Proc. 3rd Int. Conf. Early Warning (EWC)}}.
\newblock


\bibitem[\protect\citeauthoryear{{World Health Organization Media
  Centre}}{{World Health Organization Media Centre}}{2016}]%
        {who2016factsheet}
\bibfield{author}{\bibinfo{person}{{World Health Organization Media Centre}}.}
  \bibinfo{year}{2016}\natexlab{}.
\newblock \bibinfo{title}{Dengue and severe dengue}.
\newblock
  \bibinfo{howpublished}{http://www.who.int/mediacentre/factsheets/fs117/en/}.
   (\bibinfo{year}{2016}).
\newblock
\newblock
\shownote{[Online; accessed 25-April-2016].}


\end{thebibliography}
